\documentclass{article}
\usepackage[utf8]{inputenc}
\usepackage{geometry}
\usepackage{graphicx}
\usepackage{cite}
\usepackage{amsmath}
\usepackage{subfigure}
\usepackage{color}

\begin{document}
\title{Echelle Gratings with Metal Reflectors in Generic Thick Silicon Technology}

\author{Jos\'e~David~Dom\'enech, Roc\'io Ba\~nos and Pascual~Mu\~noz~
\thanks{J.D. Domenech and P. Mu\~noz are with VLC Photonics S.L., C/ Camino de Vera s/n, Valencia 46022, Spain e-mail: david.domenech@vlcphotonics.com}
\thanks{R. Ba\~nos and P. Mu\~noz are with the Optical and Quantum Communications Group, iTEAM Research Institute, Universitat Polit\`ecnica de Val\`encia, C/ Camino de Vera s/n, Valencia 46022, Spain e-mail: pmunoz@iteam.upv.es.}
\thanks{Corresponding author: P. Mu\~noz, e-mail: pmunoz@iteam.upv.es.}
}

\markboth{Journal of Lightwave Technology,~Vol.~xx, No.~yy, Month~2015}%
{Author \MakeLowercase{\textit{et al.}}: Echelle Gratings with Metal Reflectors ...}

\maketitle

\begin{abstract}
In this paper, the experimental demonstration of Echelle Grating multiplexers in generic thick Silicon technology, using metal reflectors, is reported. Two multiplexer designs are shown, covering the C-band and partially the S and L bands, with 4 and 8 channels respectively. The multiplexers exhibited performance similar to previously reported devices on dedicated manufacturing processes. The average figures measured are insertion loss 5~dB, loss non-uniformity 3~dB, polarization dependent loss 0.6~dB,  polarization dependent wavelength shift of 0.3~nm, with reduced footprint. The performance comparison between multiplexers with and without metal mirrors, for both polarizations, is provided. Several dies were measured, and the passband features for the multiplexers are analyzed, giving a reference on the process variations for future designers.
\end{abstract}


\section{Introduction}
Generic photonic integration, since its inception by 2007 \cite{misc:epixnetvision}, is becoming a foundry model for photonics, similar to what started to happen in electronics more than three decades ago at the US with the MOSIS service\cite{misc:mosis}. MOSIS provides a Multi-Project Wafer (MPW) where users share the area of wafer, and fabrication costs. Being the technology not devoted to specific purposes, that is, being generic, it may well serve different application domains, while the manufacturing is done in a common process. Hence, users benefit from MPW in terms of cost-effective prototyping, but using a fabrication process which may be up-scaled if required. 

Several generic integration platforms have emerged for different photonic technologies, or material systems, and operate nowadays \cite{munoz_icton2013}. Very recently the Silicon-on-Insulator (SOI) platform ePIXfab \cite{misc:epixfab}, which traditionally provided Silicon technology with nanometric waveguides, of typical cross-section WxH~450x220~nm$^2$, has incorporated micrometric size Silicon waveguides \cite{ipc:vtt_pw14}, despite this thick Silicon technology is well known and exists since long time ago \cite{art:soref93}. In fact, thick Si is the core technology, of several market players since 2006, but using proprietary processes \cite{rep:yole}.

Demonstrating the versatility of the generic process is key, and this is achieved by developing different photonic building blocks, a summary list of them can be found in \cite{munoz_icton2013} per platform. Thin Si photonics \cite{ibk:bogaerts14}, Indium Phosphide \cite{art:smit_iop14} and Silicon Nitride \cite{ipc:heideman09} have demonstrated a wide list of building blocks, ranging from passive waveguide and coupling components, through filters to active blocks as modulators, amplifiers, lasers and detectors. In generic thick Si technology the basic waveguiding and coupling components have been demonstrated \cite{ipc:aalto_2013}. However, key devices as wavelength multiplexers have not been reported yet. It is widely accepted that on-chip silicon multiplexers are central to a huge number of applications \cite{dai2014silicon}. Among them, the Echelle Diffraction Grating, or Echelle Grating (EG) is recognized as the multiplexing block having smallest footprint and robustness in a wide number of applications \cite{art:kotura_ptl11} compared to other alternatives such as the Arrayed Waveguide Grating \cite{art:bogaerts_pj14}. In particular, in thick Si technology where the bend radius is larger than for the thin Si nanometric waveguides, the EG proves to be very convenient.

Hence, the purpose of this paper is to report on EG multiplexers in a generic thick Silicon manufacturing platform. The paper is structured as follows. Section~\ref{sec:soa} provides a general review of the most relevant works addressing EGs in SOI technology. Next, section~\ref{sec:exp} reports on the experimental outcomes of the demonstrated devices, including statistical analysis of passband features such as insertion loss, polarization dependent loss (PDL), polarization dependent wavelength shift (PDWS), bandwidth, adjacent and non-adjacent cross-talk and the comparison between devices with and without metal reflectors for both polarizations. Finally, the conclusions are presented in section~\ref{sec:conclusion}.

\begin{table*}
  \begin{center}
    \scriptsize
\rotatebox{90}{
    \begin{tabular}{|c|c|c|c|c|c|c|c|c|} \hline
      &EG 4 channels$\dagger$&EG 8 channels&\cite{art:kotura_ptl11}&\cite{art:enablence_ptl06}&\cite{art:imec_jstqe10}&\cite{art:ibm_ptl09}&\cite{art:imec_ptl08}&\cite{art:imec_jlt07}\\ \hline\hline
      Technology&Thick Si&Thick Si&Thick Si&Thick Si&Thin Si&Thin Si&Thin Si&Thin Si \\ \hline
      Waveguide&3.00x3.00 $\mu$m$^2$&3.00x3.00 $\mu$m$^2$&3.00x3.00 $\mu$m$^2$&3.00x3.00 $\mu$m$^2$&0.45x0.22 $\mu$m$^2$&0.50x0.22 $\mu$m$^2$&0.45x0.22 $\mu$m$^2$&0.45x0.22 $\mu$m$^2$ \\ \hline
      Polarization(s)&TE+TM&TE+TM&TE+TM&TE+TM&TE&TE&TE&TE \\ \hline
      Operation*&S, C, L bands&S, C, L bands&S, C, L band&S,C bands&C-Band&C-Band&C-band&C-band \\ \hline
      Grating Facet&SiO$_2$-Al/None$\dagger\dagger$&SiO$_2$-Al$\dagger\dagger$&Al (150 nm)&--&DBR&45º corner cube&DBR&None \\ \hline
      \# channels&4&8&12&2&30&--&4&4 \\ \hline
      focal length&275 $\mu$m&500 $\mu$m&--&--&--&--&--&-- \\ \hline
      spacing&20 nm&10 nm&8 nm&70 nm&3.2 nm&3.2 nm&20 nm&20 nm \\ \hline
      FSR&198 nm&113 nm&--&--&--&--&--&-- \\ \hline
      Order&8&14&--&--&--&20&--&-- \\ \hline
      Grating length&250 $\mu$m&400 $\mu$m&--&--&--&--&--&135 $\mu$m \\ \hline
      Grating period&5.425 $\mu$m&9.75 $\mu$m&--&--&--&--&--&4.35 $\mu$m\\ \hline
      \# grooves&46&41&--&--&--&38&--&31 \\ \hline
      Passband&Gauss&Gauss&Flat&Flat&Gauss&Gauss&Gauss&Gauss \\ \hline
      \multicolumn{9}{|c|}{\bfseries Results} \\ \hline
      Insertion loss&5.68$\pm$0.32 dB&4.25$\pm$0.37 dB&1.7 dB&1.8 dB&3 dB&3 dB&1.9 dB&7.5 dB \\ \hline
      Loss uniformity&3.84$\pm$0.90 dB&3.60$\pm$1.02 dB&--&--&4 dB&0.5 dB&1.7 dB&0.6 dB \\ \hline
      PDL&0.58$\pm$0.44 dB&0.58$\pm$0.45 dB&0.2 dB&0.2 dB&n.a.&n.a.&n.a.&n.a. \\ \hline
      PDWS&0.32$\pm$0.74 nm&0.35$\pm$0.48 nm&0.3 nm&0.3 nm&n.a.&n.a.&n.a.&n.a. \\ \hline
      BW 1 dB&4.18$\pm$1.12 nm&2.05$\pm$0.29 nm&5.5 nm&--&--&--&--&-- \\ \hline
      BW 3 dB&7.30$\pm$1.32 nm&3.59$\pm$0.40 nm&--&20 nm&--&--&--&-- \\ \hline
      BW 10 dB&12.54$\pm$1.20 nm&6.23$\pm$0.48 nm&--&--&--&--&--&-- \\ \hline
      BW 20 dB&16.11$\pm$1.06 nm&8.30$\pm$0.44 nm&--&--&--&--&--&-- \\ \hline
      Adjacent Xt&-34.14$\pm$2.71 dB&-32.91$\pm$3.44 dB&-25 dB&-32 dB&-15 dB&-19 dB&-25 dB&-30 dB \\ \hline
      Non-adjacent Xt&-38.33$\pm$3.23 dB&-38.15$\pm$2.71 dB&--&--&-25 dB&--&--&-- \\ \hline
      Footprint&0.12 mm$^2$&0.27 mm$^2$&15.75 mm$^2$&3.24 mm$^2$&0.5 mm$^2$&0.05 mm$^2$&0.04 mm$^2$&0.04 mm$^2$ \\ \hline
    \end{tabular}
}
    \vspace{2mm}
    \normalsize
    \caption{\label{tab} Comparison of this work with other state of the art -journal publications- (Abbreviations: PDL, Polarization Dependent Loss; PDWS, Polarization Dependent Wavelength Shift; FSR, Free Spectral Range; n.a. does not apply; -- not available;*bands may not be covered in full;$\dagger$ 4 channel devices with and without metalized grating lines fabricated for comparison, all the figures reported in this column are for metalized devices; $\dagger\dagger$SiO$_2$ 256~nm layer and 150~nm layer of Al; DBR, Distributed Bragg Reflector).}
  \end{center}
\end{table*}

\section{\label{sec:soa}State of the art and design}
A summary of devices previously reported in journals is given in Table~\ref{tab}. For completeness, the table includes both thick and thin Si waveguide devices. The former have a typical cross-section of 3x3~$\mu$m$^2$, and the latter typically 0.45x0.22~nm$^2$. Hence, the micrometric waveguide supports both TE and TM, whereas the nanometric mainly guides TE polarization with such a cross-section. The thin Si devices have been reported for the optical telecommunication C-band (1530-1565~nm) \cite{itubands}. For thick Si devices, the EG reported in \cite{art:enablence_ptl06} is an S-C band multiplexing device for passive optical networks, whereas the one demonstrated in \cite{art:kotura_ptl11} operates at S,C and L-band as well. 

Regarding the grating line, some of them report different techniques to increase the reflectivity, to compensate, up to some extent, the Fresnel reflection loss \cite{ibk:saleh} taking place between the Si guiding layer, and the cover layer, in some cases just air, in others SiO$_2$. Hence, Distributed Bragg Reflectors (DBR) \cite{art:imec_jstqe10}\cite{art:imec_ptl08}, a 45º corner cube \cite{art:ibm_ptl09} and the deposition of metals as aluminum \cite{art:kotura_ptl11} have been reported. The latter requires additional processing steps, but the achieved reflectivity may be larger than the other techniques, albeit the former are attained within the same waveguide lithography.

EGs with 4 and 8 channels (EG4 and EG8 onward) were designed according to the specifications in the first two columns of Table~\ref{tab}. The design process is as reported in \cite{art:lycett-photond}. The devices are based on a curved facet type grating and make use of input/output waveguides with the following characteristics: width 2.4~$\mu$m, side etching 1.75~$\mu$m, input waveguides center at 36$^{\circ}$ and output waveguides center at 41$^{\circ}$ and 39$^{\circ}$ for the 4 and 8 channel device respectively. The slab mode effective and group indices were computed to be 3.46 and 3.6 respectively. 

In the design area, two EG4 devices, with and without metalization, were included. All the EG8 devices include metal mirrors, as stated in Table~\ref{tab}.

\section{\label{sec:exp}Experimental results}
\subsection{Fabrication}
\begin{figure}
  {\par\centering
   \resizebox*{0.48\textwidth}{!}{\includegraphics*{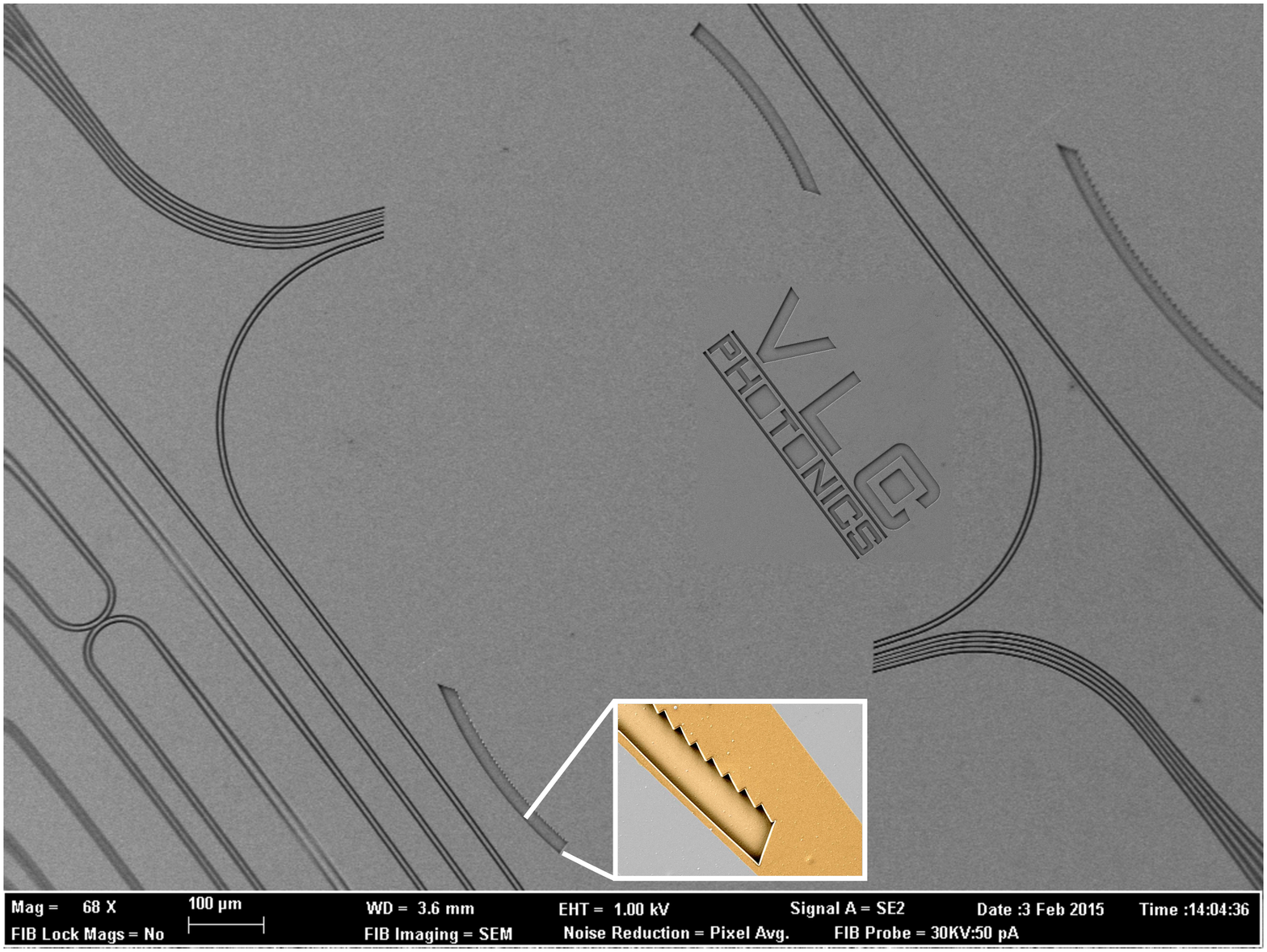}}
   \caption{FESEM micrographs of fabricated devices, EG4 device. Inset: detail of the grating line (metalization in pseudo-color).}
  \label{fig:fesem}
  }
\end{figure}

The devices were fabricated in the generic technology platform provided by VTT Finland \cite{ipc:aalto_2013}. The typical performance for 3~$\mu$m width waveguides is: propagation loss 0.1~dB/cm, fiber i/o coupling up to 90\% (lensed fibers), or 40\% (regular fiber) and small polarization dependence\cite{ipc:vtt_pw14}. Additional metalization steps are available in the platform, and they were employed to cover some of the grating lines in our devices. To be precise, the mirrors are formed by depositing two thin layers on top of the Silicon guiding layer: a first layer of SiO$_2$ (265~nm, corresponding to a quarter-wave), and a second layer of 150~nm aluminum. As reported by the manufacturers, reflectivity of these combined layers can be as good as 92.7\% \cite{art:Cherchi_OpEx_15}. Using other reflective devices within the same dies \cite{art:fandino_opex15} our conclusion is the reflectivity in these chips is approximately 79,4~\%. A Field Emission SEM (FESEM) micrograph of the fabricated devices is shown in Fig.~\ref{fig:fesem}, with an inset showing the view of the grating line, where the metal area is highlighted in pseudo-color.

\subsection{Characterization setup}
The characterization setup consists of a set of three motorized positioners. Two of them are used for holding the in/out coupling stages for end-facet chip coupling whereas the third one holds the sample on top of a thermally controlled (25~$^{\circ}$C) vacuum chuck. A microscope vision system is mounted and used for visual alignment employing red light in the in coupling stage, and a LED lamp is used for illumination. The in coupling stage consists of a fiber holder, collimating lens, free space polarizer and microscope objective. The purpose of this stage is to set the polarization of light that is injected into the chip. The light is collected out from the chip using a lensed fiber. Nontheless, in order to discard polarization rotation issues, the first measurements were done with an output stage identical to the input one. With the filters set in orthogonal polarizations, no power was collected at the ouptut. Therefore the lensed fiber was used to collect light for simplicity.

For the measurements, the in/out stages are aligned manually in two steps. Firstly, they are approximated to the waveguide locations by visual inspection using the microscope and a using red light source. The approximated distance and height are determined visually from the red spot on the edge of the chip. Secondly, without red light, a broadband light infrared source is connected to the input coupling stage, whereas a power meter is connected at the end of the other. Hence, the positions of the input microscope objective and the output fiber are optimized with the motorized stages to obtain the maximum power. After the alignment is optimized, an Optical Spectrum Analyzer (OSA) is used to record the infrared spectra with a resolution of 10~pm.

\begin{figure}
  {\par\centering
   \subfigure[EG4]{\resizebox*{0.48\textwidth}{!}{\includegraphics*{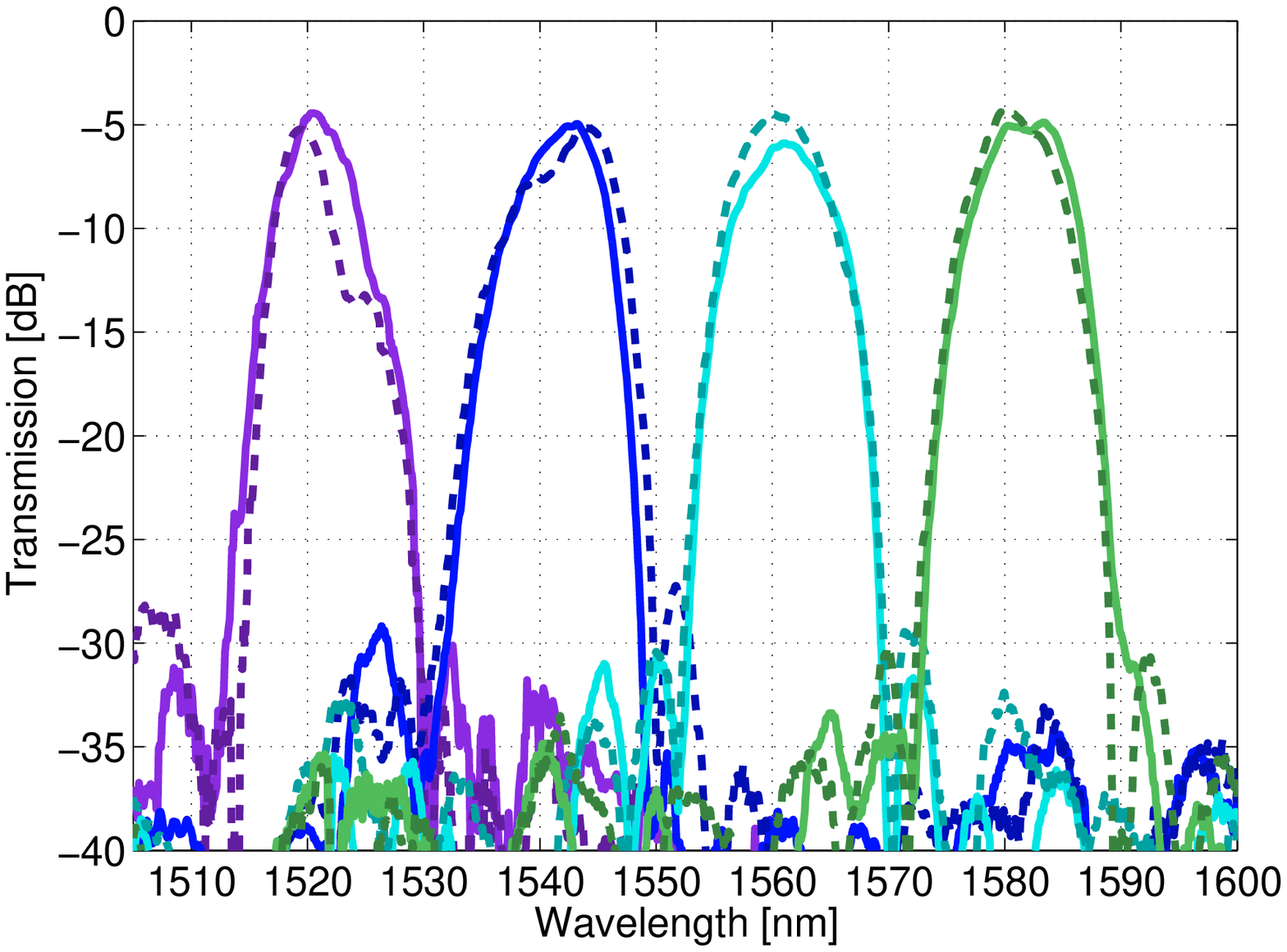}}}
   \subfigure[EG8]{\resizebox*{0.48\textwidth}{!}{\includegraphics*{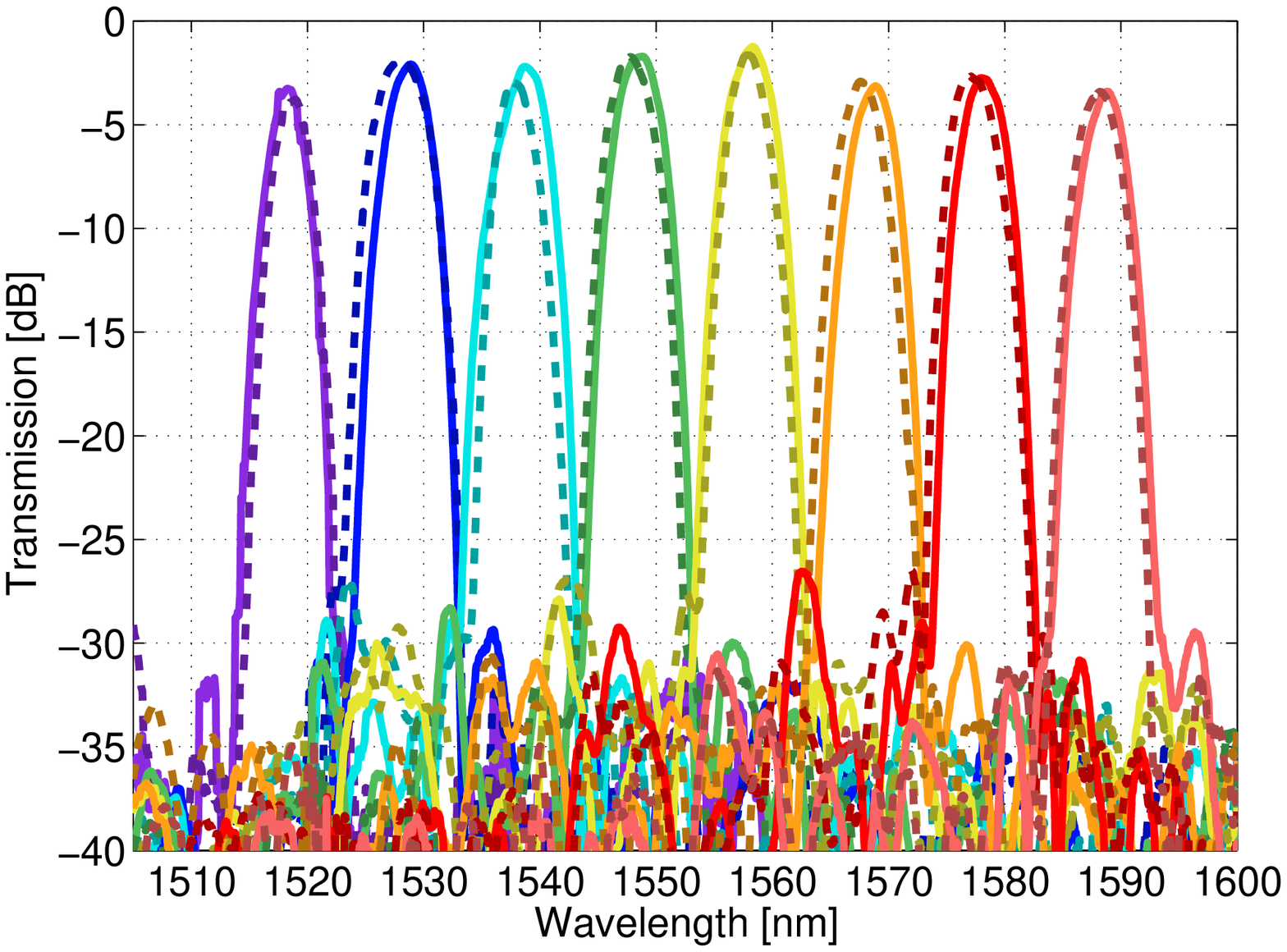}}}
   \caption{Transmission spectra for the EG with metal reflectors, 4 channel (a) and the 8 channel EG (b) devices. Solid and dashed lines correspond to TE and TM polarization respectively.}
  \label{fig:spectra}
  }
\end{figure}

\subsection{Transmission Spectra}
The transmission spectra for two samples of EG4 and EG8 devices are shown in Fig.~\ref{fig:spectra}-(a) and (b) respectively. The results are normalized with respect to a straight (test structure) waveguide. The minimum insertion loss for the EG4 is around 5~dB, versus 2~dB for the EG8 device. Conversely, the EG8 exhibits higher loss non uniformity, around 2~dB, compared to 1~dB for the EG4 design. This is in agreement with the channel spread within the free spectral range, which is larger for the EG8 (80~nm within the FSR of 113~nm) compared to the EG4 design (80~nm within an FSR of 198~nm), which happens to be a common feature with other diffractive multiplexers \cite{art:munoz_jlt02}. 

The graphs also show the overlaid spectra for TE and TM polarizations, which were set using bulk filters in the in coupling stage mentioned previously. In accordance to the small polarization dependence of this thick Si technology, from the graphs low PDL and PDWS are observed. For these particular devices PDL is as low as 0.2~dB, except for the third channel in EG4 on Fig.~\ref{fig:spectra}-(a), and PDWS is as low as 0.3~nm. Both values are in good agreement with other reported thick Si EGs \cite{art:kotura_ptl11}\cite{art:enablence_ptl06}.

Finally, and less noticeable, all the passbands have a common feature, that is a small side lobe close to the left side of the main lobe (towards shorter wavelengths). This side lobe is also predicted in the design and simulation stage, where slanted grating sidewall (between 1-2$^{\circ}$) were included. The slanted grating walls result into slightly increased insertion loss in the main band, due to the coupling to higher order (vertical) slab modes in the EG, which in turn result into the aforementioned left side lobe present in all the spectral traces \cite{art:kotura_ptl11}.

\begin{figure}
  {\par\centering
    \resizebox*{0.48\textwidth}{!}{\includegraphics*{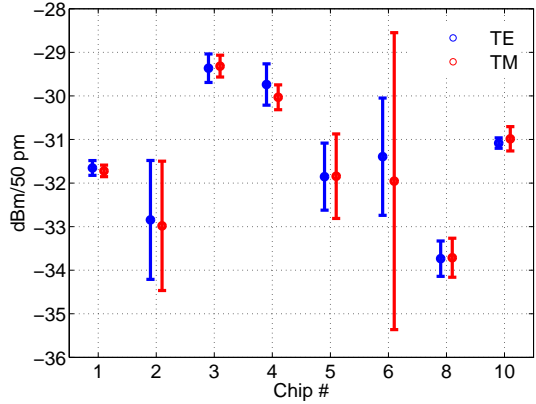}}
    \caption{Average transmission at 1550~nm, and standard deviation, on test straight waveguides per chip, for TE (blue) and TM (red) polarizations.}
    \label{fig:coupling}
  }
\end{figure}

\subsection{Passband features analysis}
Several dies were retained for full measurements, having as target the analysis of variations among them. Each die includes test structures and four EGs, two of them are the EG4 design with and without metal mirrors. The other two are EG8 designs. In some cases the results shown onwards are only for TE polarization for the sake of clarity, since for TM similar results were obtained. EG devices from five dies were measured in full, and other dies were discarded due to randomly damaged i/o waveguide facets for the multiplexers. 

In fact, this lead to investigate the possibly different coupling levels to the waveguides. Hence, Fig.~\ref{fig:coupling} shows the mean transmitted power, and standard deviation, at 1550~nm, derived from sets of four test straight waveguides per die, and for the two polarizations. The overall average for all the waveguides and chips is -31.35~dBm/50~pm and -31.4628~dBm/50~pm for TE and TM respectively, with corresponding standard deviations of 1.55~dB and 1.86~dB. These results are relevant for the evaluation of all the pass band features related to power, that are presented in the following paragraphs, such as insertion loss, loss non-uniformity and PDL).

\begin{figure}
  {\par\centering
   \subfigure[EG4]{\resizebox*{0.48\textwidth}{!}{\includegraphics*{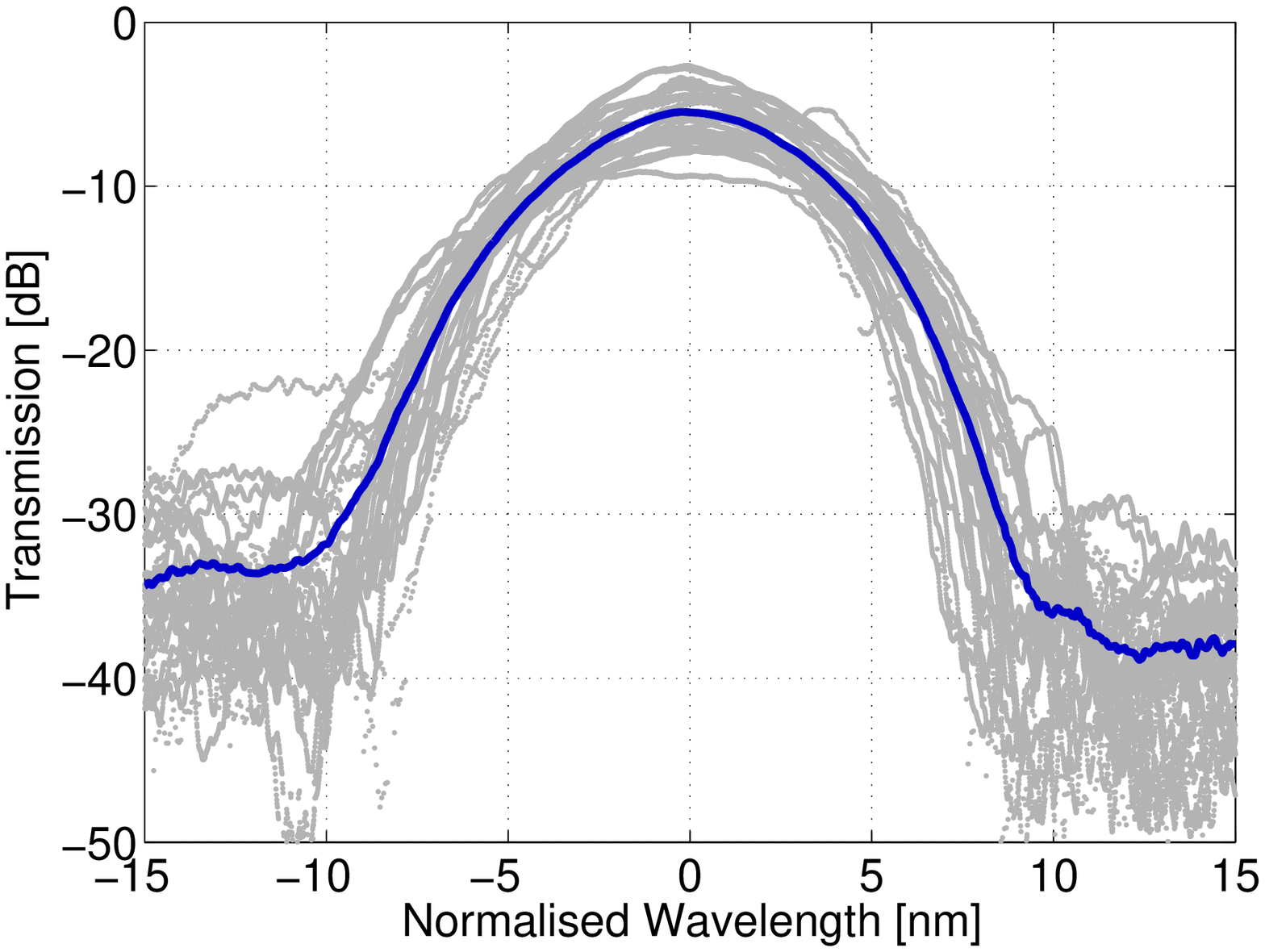}}}
   \subfigure[EG8]{\resizebox*{0.48\textwidth}{!}{\includegraphics*{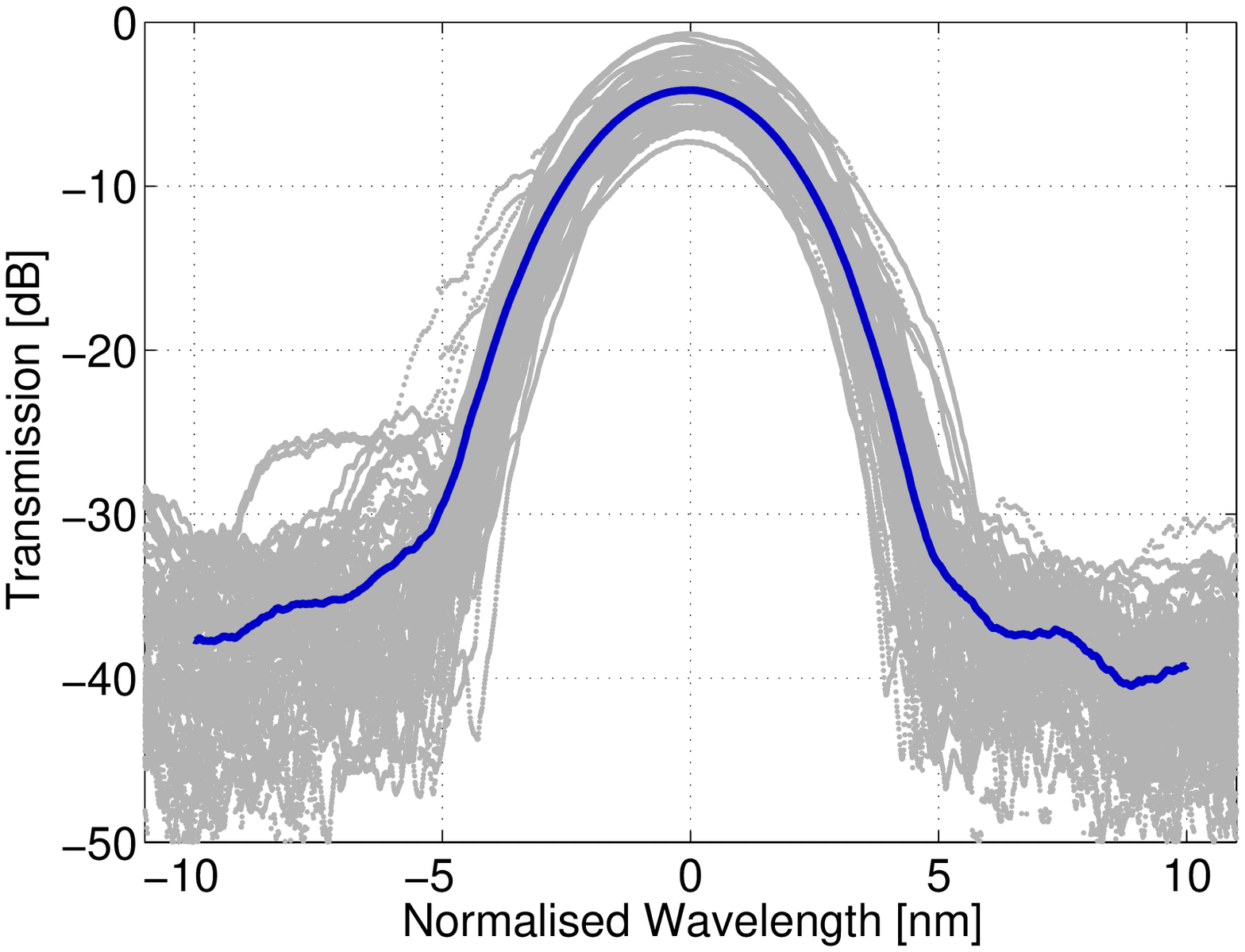}}}
   \caption{Superimposed transmission spectra for the 4 channel (a) and 8 channel EG (b) devices. Grey dots show the overlaid spectra for all measured channels from 5 chips, 1 device per chip. All the channels have been aligned to the same center wavelength. The blue line shows the average response. TE polarization, TM polarization exhibited similar performance.}
  \label{fig:cloud}
  }
\end{figure}

Fig.~\ref{fig:cloud} shows the superimposed tranmission spectra for all the channels, aligned to the (normalized) wavelength corresponding to the passband 3-dB bandwidth center, both for EG4 and EG8, in panels (a) and (b) respectively. The collected statistical features from the passbands are summarized in the first two columns of Table~\ref{tab}. As mentioned previously, note that a part of all the variations reported for the figures related to optical power can be attributed to different facet condition for the different EG input/output waveguides, and that they may be reduced by facet polishing \cite{art:bowers_jlt14}.

In terms of channel insertion loss, the average of the peak insertion loss for all channels and dies was calculated. EG4 and EG8 exhibit 5.7~dB and 4.3~dB with little variations, as detailed in the table. Compared to previous reported values, Table~\ref{tab}\cite{art:kotura_ptl11}\cite{art:enablence_ptl06}, these are 2-4~dB larger. This maybe attributed to a non-optimized generic process which was also used for the first time to this purpose, compared to the industrial dedicated process from \cite{art:kotura_ptl11}\cite{art:enablence_ptl06}. Next, the loss uniformity of the devices was calculated as the difference between the maximum and minimum peak transmission for each device. The mean and standard deviation over all the dies is also shown in Table~\ref{tab}, where similar results were obtained for EG4 and EG8, that is an average uniformity around 3.7~dB with variations of around $\pm$1~dB. 


The PDL was calculated as the difference of the peak amplitude for TE and TM polarization of a given channel. The results are similar for EG4 and EG8, where the average PDL is close to 0.6~dB, whereas other thick Si references, Table~\ref{tab}\cite{art:kotura_ptl11}\cite{art:enablence_ptl06} reported a PDL of 0.2~dB. In terms of PDWS, computed as the difference between the wavelengths were the TE and TM maxima per channel are locateld, our results are on average close to those previously reported of 0.3~nm.

\begin{figure}
  {\par\centering
   \resizebox*{0.48\textwidth}{!}{\includegraphics*{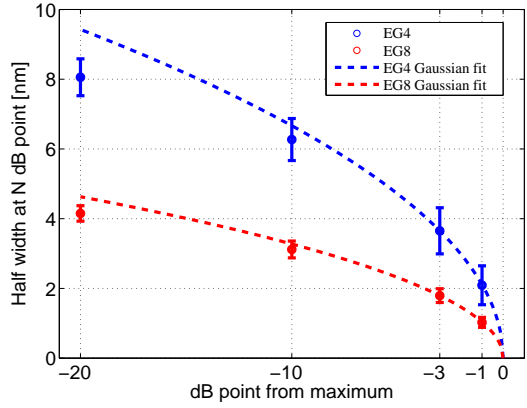}}
   \caption{Measured averaged N-dB bandwidth points and Gaussian fit for the 4 channel (blue) and 8 channel (red) EGs. TE polarization, TM polarization exhibited similar behavior.}
   \label{fig:bw}
  }
\end{figure}

The bandwidth was calculated from the measurements for all the devices and channels, and the obtained bandwidth values were averaged over all the channels and samples as well. The N-dB bandwidth was calculated as the wavelength difference between the two poins in the passband N-dB under the maximum. The average and standard deviation interval for the EG4 and EG8 devices are depicted in Fig.~\ref{fig:bw}. A Gaussian power transfer function was fitted to the 3-dB bandwidth, and is shown in dashed lines within the same graph. Both devices exhibit a good match to a Gaussian band pass down the 10-dB point from the maximum. This is also the case for the EG8 down to the 20-dB point. However, the EG4 device exhibits larger deviation from a Gaussian response at the 20-dB point, compared to the EG8 design.


\begin{figure}
  {\par\centering
    \subfigure[]{\resizebox*{0.48\textwidth}{!}{\includegraphics*{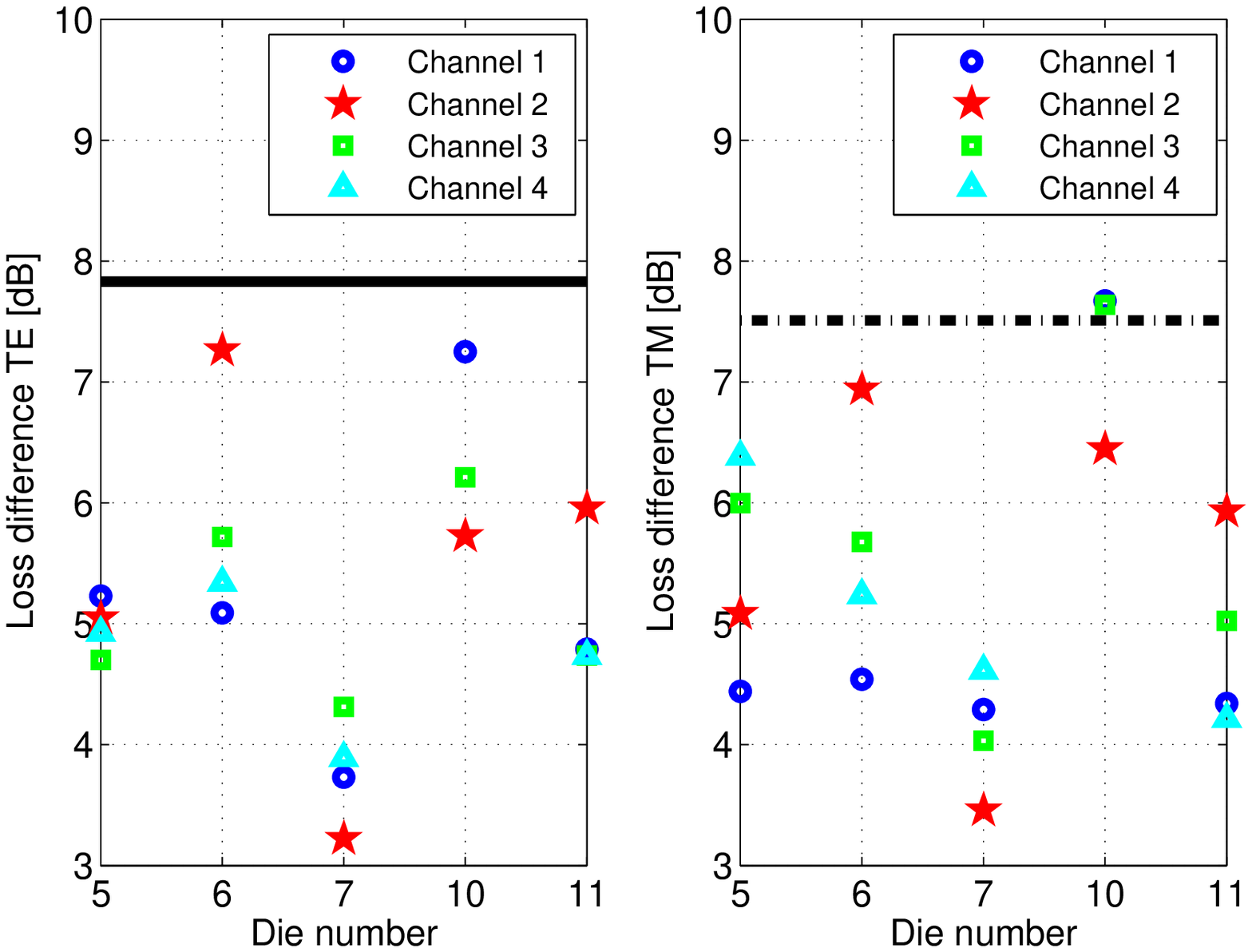}}}
    \subfigure[]{\resizebox*{0.48\textwidth}{!}{\includegraphics*{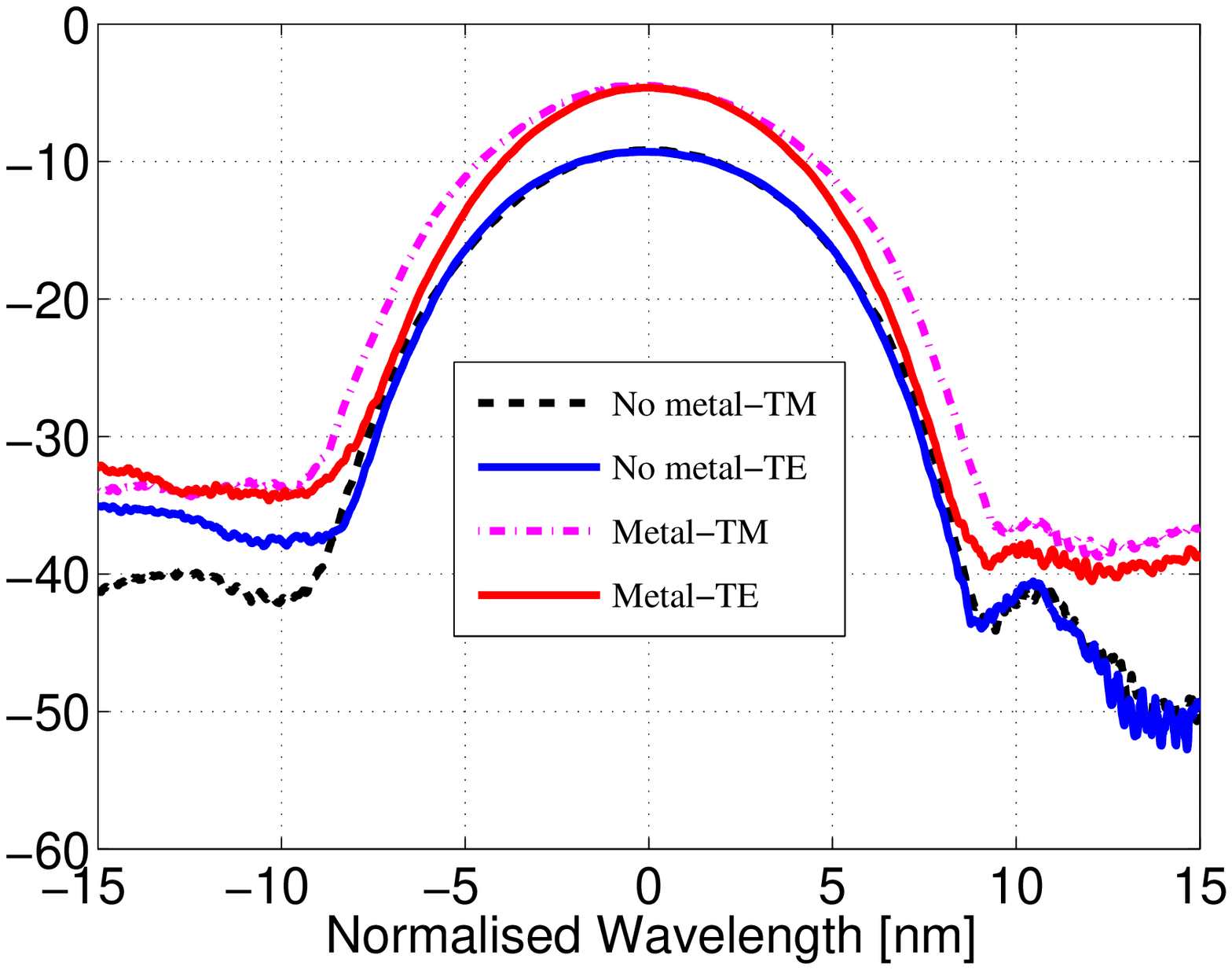}}}
   \caption{Loss difference between the metalized and non metalized grating line devices, EG4 different dies (5,6,7,10,11) in (a), where each EG channel is represented with different symbol (legend), and the Fresnel reflection loss limit is marked in a black line, for TE and TM. Comparison of averaged responses from all the EG4 channels (b), with and without grating line metalization, and for TE and TM polarizations.}
  \label{fig:loss}
  }
\end{figure}

The cross-talk was computed as the maximum value of the adjacent (non-adjacent) channel(s) in a given channel 3-dB clear window \cite{art:bowers_jlt14}. The average for adjacent channel cross-talk is around -33~dB and -34~dB for the EG4 and EG8 respectively. This is slightly better than reference \cite{art:kotura_ptl11} and similar to reference \cite{art:enablence_ptl06}. The non-adjacent channel cross-talk is approximately -38~dB for both devices. 

\subsection{Metal mirrors analysis}
Fig.~\ref{fig:loss}-(a) shows the loss difference per channel between EG4 devices with and without metal mirrors, and for both polarizations. Excluding die number 7, the difference varies between 5 and 7~dB approximately both for TE and TM. Fig.~\ref{fig:loss}-(b) shows the averaged spectra for all channels, analog to Fig.~\ref{fig:cloud}. The difference for the EG4 metalized and non-metalized devices is 4.62~dB and 4.67~dB for TE and TM. In all the cases, both devices (with and without mirror) sit on the same die, that is the grating line patterning is expected to be similar, despite the orientation is different, as can be seen in Fig.~\ref{fig:fesem}. 

In an ideal situation, one would expect the loss difference between devices with and without mirrors to be close to the Fresnel reflection loss. Since the difference between the incident (input waveguide) and diffraction (output waveguides) angles is 5$^{\circ}$, the Fresnel reflection loss contribution is calculated to be 7.83~dB for TE (for SiO$_2$ cover, for air it is around 4.7~dB \cite{art:imec_jlt07}), and 7.51~dB for TM. Using the mirror structure described above, ideally the loss in the interface would 0.069~dB and 0.064~dB for TE and TM respectively, slightly less lossy for TM \cite{ibk:orfanidis}. Nonetheless we measured these SiO$_2$-Al mirrors reflectors in the dies provide a reflectivity in average of 79.4~\% (1.0~dB loss) \cite{art:fandino_opex15}. 

Hence, this approximately 2.0~dB variations in agreement the ones described from Fig.~\ref{fig:loss}-(a), which may be attributed, besides to poorer performance of the mirrors in some cases, to different sources of errors: different imperfections in the grating lines of metalized and non metalized devices (roughness, rounded corners, ...) as well as differences in the in/out coupling properties of the different samples used to average and substract. 
\section{\label{sec:conclusion}Conclusion and outlook}
In this paper we reported on Echelle Gratings multiplexers designed and manufactured using thick Silicon photonic integration technology. The paper supplies information of variations on different figures of merit for the multiplexers. Furthermore, the comparison between devices with and without metal mirrors is supplied as well, for different polarizations. 

\section*{Acknowledgment}
The authors acknowledge financial support by the Spanish CDTI NEOTEC start-up programme, the Spanish MINECO project TEC2013-42332-P, acronym PIC4ESP, project FEDER UPVOV 10-3E-492 and project FEDER UPVOV 08-3E-008. The authors acknowledge J.S. Fandi\~no for helpful discussions.



\bibliographystyle{IEEEtran}
\bibliography{vtt_eg_jlt}

\end{document}